\documentclass[%
groupedaddress,
twocolumn,
amsmath,amssymb,
aps,
prl,
]{revtex4-1}

\usepackage{graphicx,dcolumn,amsmath,amssymb,color,mathrsfs,titlesec,hyperref,enumerate,bm,subfigure,float}
\definecolor{linkColor}{rgb}{1,0,0}
\hypersetup{pdfborder={0 0 0},colorlinks=true,urlcolor=linkColor,citecolor=linkColor}

\usepackage{mathtools}
\DeclarePairedDelimiter{\ceil}{\lceil}{\rceil}
\DeclarePairedDelimiter{\floor}{\lfloor}{\rfloor}
\begin{document}
\preprint{APS/Code}

\title{Large-Scale Experimental and Theoretical \\ Study of Graphene Grain Boundary Structures}

\author{Colin Ophus}
\email[]{cophus@gmail.com}
\affiliation{National Center for Electron Microscopy, Molecular Foundry, Lawrence Berkeley National Laboratory, Berkeley, CA, USA}

\author{Ashivni Shekhawat}
\email[]{shekhawat.ashivni@gmail.com}
\affiliation{Miller Institute for Basic Research in Science, Berkeley, CA, USA}
\affiliation{Materials Science Division, Lawrence Berkeley National Laboratory, Berkeley, CA, USA}

\author{Haider Rasool}
\affiliation{Department of Physics, University of California at Berkeley, Berkeley, Ca, USA}
\affiliation{Materials Science Division, Lawrence Berkeley National Laboratory, Berkeley, Ca, USA}

\author{Alex Zettl}
\affiliation{Department of Physics, University of California at Berkeley, Berkeley, Ca, USA}
\affiliation{Materials Science Division, Lawrence Berkeley National Laboratory, Berkeley, Ca, USA}

\date{\today} 

\begin{abstract}
We have characterized the structure of 176 different single-layer graphene grain boundaries using  $>$1000 experimental HRTEM images using a semi-automated structure processing routine. We introduce a new algorithm for generating grain boundary structures for a class of hexagonal 2D materials and use this algorithm and molecular dynamics to simulate the structure of $>$79~000 graphene grain boundaries covering 4122 unique orientations distributed over the entire parameter space. The dislocation content and structural properties are extracted from all experimental and simulated boundaries, and various trends are explored.  We find excellent agreement between the simulated and experimentally observed grain boundaries. Our analysis demonstrates the power of a statistically significant number of measurements as opposed to a small number of observations in atomic science. All experimental and simulated boundary structures are available online.
\end{abstract}
\pacs{PACS numbers}
\maketitle

\section{Introduction}

Single-layer graphene is a promising material for many technological applications, due to its excellent mechanical~\cite{lee2008,wei2012nature, rasool2013measurement} and electronic properties~\cite{li2009large, neto2009electronic}. Most graphene deposition methods  produce polycrystalline sheets, containing grain boundaries (GBs) \cite{li2009large, rasool2010continuity}.  This polycrystal structure introduces local property deviations at the boundaries that could limit or enable various potential applications.  There is also strong scientific interest in graphene GBs due to their one-dimensional nature.  Some examples include a bimodal phonon scattering behaviour across graphene GBs \cite{yasaei2015bimodal}, anomalous strength characteristics \cite{grantab2010anomalous, rasool2013measurement}, strong chemical sensitivity of boundary charge transform~\cite{yasaei2014chemical}, a transformation  of the GBs from transparency of charge carriers to near-perfect reflection \cite{yazyev2010electronic}, amongst others. 

A large number of theoretical studies on graphene GB structures have been performed by many researchers \cite{yazyev2010electronic, yazyev2010topological, liu2010cones, malola2010structural, cockayne2011grain, carlsson2011theory, yi2013theoretical, zhang2013structures, tan2013effect, dai2014electronic, vancso2014effect, grantab2010, zhang2012, yakobson2013, liu2012, hao2012, shenoy2014, gao2012, yannik2012, jannik2013, cao2013}.  However the number of corresponding experimental measurements of free-standing graphene GB structure at atomic resolution is much smaller \cite{huang2011grains, an2011domain, kim2011grain, kurasch2012atom, rasool2013measurement, rasool2014conserved}.  These experimental studies have typically considered either a single boundary structure, or a small number of GB structures. Thus, the gap between the number of possible or proposed graphene GB structures and those experimentally observed is very large. This makes testing the various proposed models for graphene GB structure and structure formation very challenging \cite{yazyev2010topological, guo2015governing}.



In this study, we have characterized the structure $\approx$176 graphene GB structures from 1067 phase-contrast high resolution transmission electron microscopy (HRTEM) observations of free-standing single-layer graphene samples. We have characterized the atomic structure  using a semi-automated processing routine, measuring the local topology and various other physical parameters.  We have also used a new algorithm to generate the structure of $\approx$79,000 graphene GBs covering the entire orientation parameter space, which were then relaxed using molecular dynamics (MD).  We have performed a detailed structural characterization of all experimental and simulated boundaries, extracting structure parameters and dislocation content of all boundaries.  The proposed algorithm for generating graphene GB structures is found to be in excellent agreement with the observed structures.


\section{Methods: Experimental}
\subsection{Graphene Sample Fabrication and HRTEM Imaging}

Graphene samples were grown on polycrystalline copper substrates at 1035$^\circ$C by chemical vapor deposition. The substrate was held at 150 mTorr hydrogen for 1.5 hours in a quartz tube furnace, and then 400 mTorr methane was run (5 sccm) over the sample for 15 minutes to grow single-layer graphene. Further details of this method are described in Refs.~\cite{li2009large, rasool2010continuity, rasool2013measurement}.  


All experimental high-resolution transmission electron microscope (HRTEM) images used in this study were recorded on the TEAM 0.5, a monochromated, aberration-corrected FEI Titan-class microscope, operated at 80 kV with a high brightness Schottky field emission gun. Rather than optimizing the imaging conditions, we instead focused on recording images as quickly as possible so as to minimize electron beam damage of the GBs. Often, multiple images of the same boundary were collected sequentially which allowed for both optimization of the imaging conditions and observations of beam-induced modifications of the structure.

\subsection{Semi-Automated Experimental GB Analysis}

\begin{figure}[htbp]
    \centering
        \includegraphics[width=3.4in]{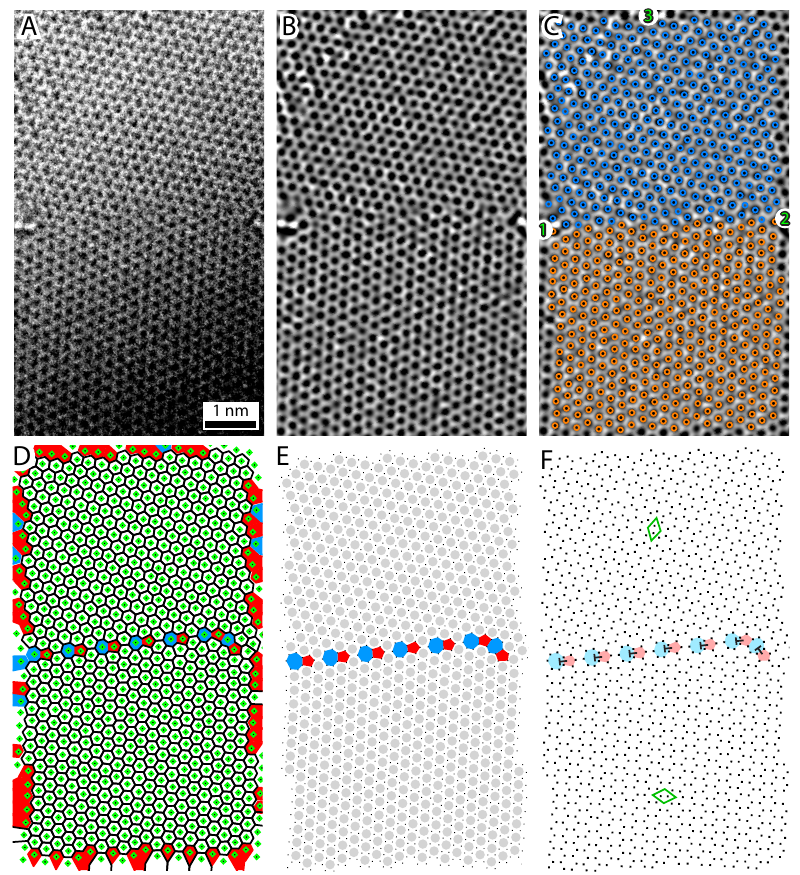}
 	\caption{Example of GB structure parsing for an experimental image: {\bf(A)} original micrograph, {\bf(B)} local intensity-ordered filtered image, {\bf(C)} ring center detection and user-selected boundaries, {\bf(D)} Voronoi tessellation, {\bf(E)} edge cleanup and atomic coordinate generation, and {\bf(F)} final coordinates with dislocation identification.}
	\label{FigureExpParse}
\end{figure}

The graphene HRTEM images have a low signal-to-noise ratio due to the low scattering efficiency of single carbon atoms. In order to measure the boundary structure for hundreds of images, we have developed a processing routine with as few user inputs as possible. This routine is shown schematically in Fig.\ \ref{FigureExpParse}. First, a linear background is fitted and removed from the image to minimize the intensity falloff caused by the monochromatic aperture. Next, a nonlinear filter is applied to remove noise (rank filter of local intensity values within a 0.6\AA~radius, 25$\%$ darkest value selected \cite{soille2002morphological}), shown in Fig.\ \ref{FigureExpParse}B. Peak detection is used to find local minima, and the user inputs the boundary extent, the three locations labeled in Fig.\ \ref{FigureExpParse}C. A subset of the detected peaks given by this user-selected region-of-interest is used to automatically characterize the boundary structure.

The first step of the boundary characterization is a Voronoi tessellation of the detected local minima, shown in Fig.\ \ref{FigureExpParse}D. The Voronoi cell vertices represent atomic positions. The number of carbon atoms in each ring is given by the number of sides of each cell. Next, the boundary cells are removed and the tessellation is recomputed using a weighted Voronoi algorithm \cite{aurenhammer1984optimal}, with the weights set to keep the mean bond length constant for all cells, shown in Fig.\ \ref{FigureExpParse}E. The final atomic coordinates are plotted in Fig.\ \ref{FigureExpParse}F, with a final optimization performed to enforce a minimum distance constraint of 1.2~\AA\ on all atomic coordinates, to ensure a physically realistic result.  The boundary can be traced by connecting all pentagon and heptagon rings, and a best-fit lattice is calculated for the two grains on each side.  The dislocations are located by searching for a minimal description of all pentagon-heptagon pairs.

\begin{figure}[htbp]
    \centering
        \includegraphics[width=3.4in]{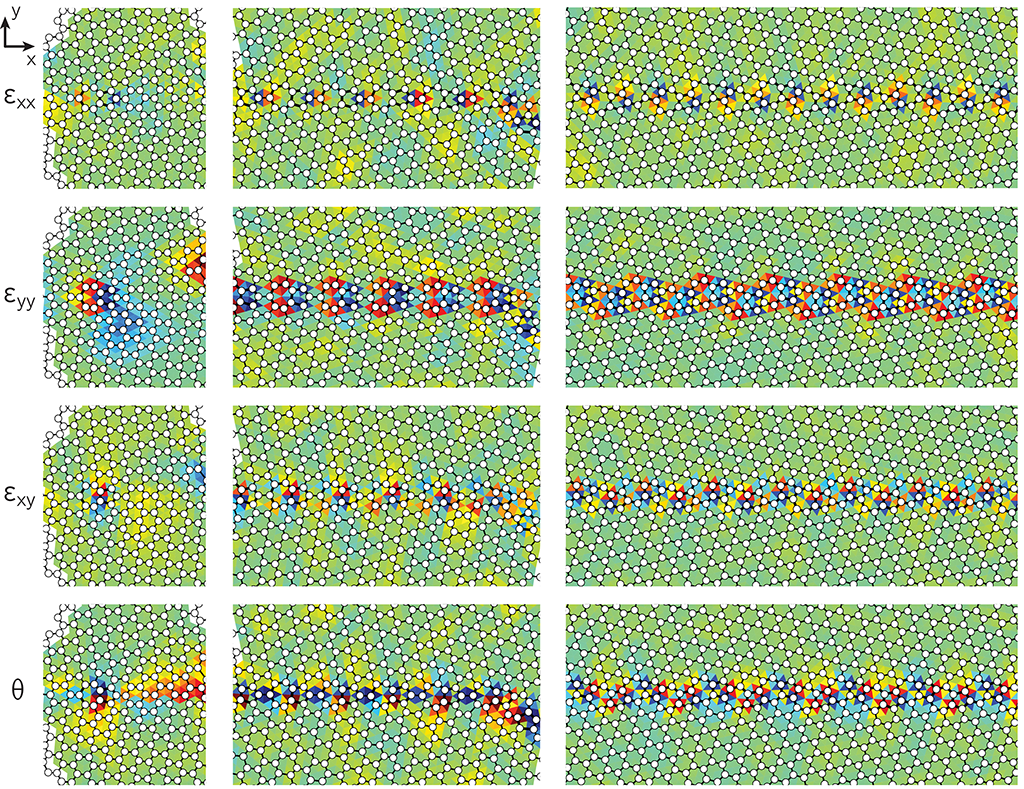}
 	\caption{Examples of atomic strain measurements on three experimentally measured boundaries. The values of $\epsilon_{\rm{xx}}$, $\epsilon_{\rm{yy}}$, and $\epsilon_{\rm{xy}}$ range from -25\% (blue) to +25\% (red), while local rotation $\theta$ has a range of $\pm 20^\circ$ from blue to red.}
	\label{FigureExpStrain}
\end{figure}

Additionally, the strain of the experimental boundaries was estimated using a geometric method similar to that given in Ref.~\cite{mott1992atomic}. Each atom is placed at the center of a triangle defined by its 3 nearest-neighbors, calculated from a Delaunay triangulation of the set of atoms.  These triangles are referenced to the appropriate triangle (2 atomic sites per unit cell) formed by the lattice vectors of the best-fit lattices of the two grains.  The linear transformation matrix for each triangle is used to calculate local strains (infinitesimal strain is assumed for convenience). Rather than parse the strain into atomic coordinates as in Ref.~\cite{mott1992atomic}, we instead calculate the root-mean-square values of the local strains and the local rotation, since we are interested in the ``average'' distortion of each of the boundaries.  Three examples of these strain measurements are given in Fig.~\ref{FigureExpStrain}.

\section{Methods: Numerical}

\subsection{Space of Graphene GBs}

\begin{figure}[htbp]
\begin{center}
\includegraphics[width=3.4in]{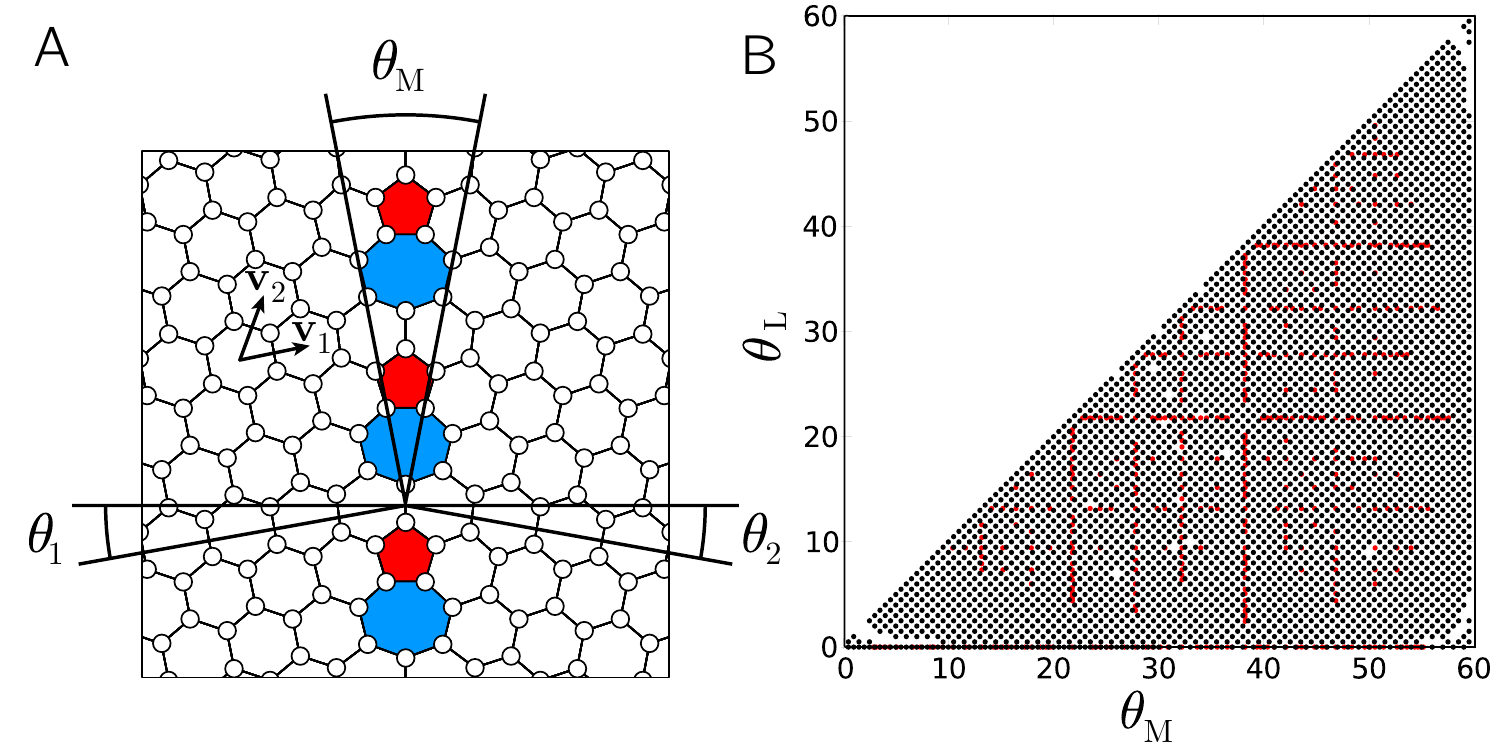}
\caption{(Color online) {\bf (A)} A typical grain boundary structure showing the graphene lattice vectors $\mathbf{v}_{1}$ and $\mathbf{v}_{2}$, and the definition for the grain angles $(\theta_{1},\theta_{2})$. The misorientation and the line angles are defined as $\theta_{\rm{M}} = \theta_1 + \theta_2$, and $\theta_{\rm{L}} = |\theta_1 - \theta_2|$. {\bf (B)} A representation of all the $(\theta_{\rm{M}}, \theta_{\rm{L}})$ pairs simulated in this study. The red dots indicate the perfectly commensurate GBs, while the black dots indicate the approximately commensurate ones. A total of 4122 $(\theta_{\rm{M}}, \theta_{\rm{L}})$ pairs were simulated, resulting in over 79,000 GB structures when including the displacement $\Delta$.}
\label{fig:gbspace}
\end{center}
\end{figure}

Bulk three-dimensional materials require 5 angles to characterize the macroscopic degrees of freedom of a general GB, while two-dimensional materials require only 2 angles. Thus the parameter space for 2D GBs is far smaller than that of 3D grain boundaries.
As shown in Fig.~\ref{fig:gbspace} A these two angles are the misorientation angle $\theta_{\rm{M}} = \theta_1 + \theta_2$, defined as the angle between the unit cell vectors of each grain, and the boundary line direction $\theta_{\rm{L}} = |\theta_1 - \theta_2|$, defined as the angle between the boundary vector and the symmetric tilt boundary vector for a given $\theta_{\rm{M}}$. Due to the symmetries of the graphene lattice we get $\theta_{\rm{M}} \in (0^\circ, 60^\circ)$ and $\theta_{\rm{L}} \in (0^\circ,\theta_{\rm{M}})$. A third parameter, namely the relatively sliding of the two grains along the GB boundary is also needed for a complete description of the boundary. In our simulations we choose the relative sliding that gives the lowest GB energy, thus effectively eliminating this degree of freedom. 

In order to minimize the boundary effects, we simulate GB structures that are periodic along the GB direction. This requirement places strong restrictions on the GB configurations that we can simulate. Consider simulation of the GB corresponding to a point $(\theta_{\rm{M}},\theta_{\rm{L}})$, or equivalently $(\theta_1,\ \theta_2)$, in the parameter space. The lattice vectors for graphene are $\mathbf{v}_{1} = a\sqrt{3}\mathbf{e}_1$, $\mathbf{v}_{2} = a\sqrt{3}/2\mathbf{e}_1 + 3a/2 \mathbf{e}_2$, where $a$ is the carbon-carbon bond length, and $\mathbf{e}_{1,2}$ are unit vectors parallel and perpendicular to the zigzag axis of the graphene sheet, respectively. Thus for given $(\theta_1,\ \theta_2)$ the corresponding grains have periodic repeat distances of $l_i = a\sqrt{3(n_{i1}^2 + n_{i2}^2 + n_{i1}n_{i2})}$ ($i = 1,\ 2$) along the GB direction, where $n_{i1},\ n_{i2}$ are integers such that $\tan\theta_i = (2n_{i1} + n_{i2})/\sqrt{3} n_{i2}$~\cite{saito1998physical}. For an arbitrary $\theta_i$ there may exist no suitable integers $n_{i1},\ n_{i2}$, or even if such integers exist, $l_i$ can be prohibitively large for MD simulation. Further, in order to simulate a GB, $l_1$ and $l_2$ should either be commensurate, i.e., $l_1/l_2 = p/q$ where $p,\ q$ are positive integers, in which case the net GB length is given by $l_{GB} = ql_1 = pl_2$, or they should be approximately commensurate, i.e., $l_1/l_2 \approx p/q$, in which case the simulated GB length is $l_{GB} = 2l_1ql_2p/(l_1q + l_2p)$, and each grain has a strain of magnitude $|l_1q - l_2p|/(l_1q + l_2p)$. In case of approximately commensurate boundaries, we require that the rational approximation $p/q$ is chosen such that the resulting strain magnitude is less than $10^{-4}$, so that the resulting elastic distortion is minimal. Due to numerical considerations, we simulated GBs with a maximum length of 2000\ \AA. If for a given $(\theta_1,\ \theta_2)$ the GB length is greater than 2000\ \AA, we try to find a nearby GB such that the resulting grain angles are within 0.01$^\circ$ of the desired values. Finally, in order to choose the relative sliding between the two grains that leads to minimum GB energy, we search in steps of 1 \AA\ over the entire range $\Delta$~\cite{sethna2008}, given by
\begin{equation}
\Delta = \left\{\begin{array}{ll}
    l_1 & \rm{ if\ commensurate} \\
    \min\Big( |l_2  - \floor{l_2/l_1}l_1|, & \rm{if\ approximately} \\
    |l_2 - \ceil{l_2/l_1}l_1|\Big), &  \rm{\phantom{if}commensurate}
\end{array}\right.
\end{equation}
assuming $l_1 < l_2$. We simulate all perfectly commensurate GBs with length less than 2000 \AA\ , and grid the $(\theta_1,\ \theta_2)$ space in steps of $0.5^\circ$, resulting in a `diagonal gridding' of the $(\theta_{\rm{M}},\ \theta_{\rm{L}})$ space in steps of $1.0^\circ$. However, for certain configurations near $(\theta_{\rm{M}},\ \theta_{\rm{L}})$ $=(0^\circ,\ 0^\circ),\ (60^\circ,\ 0^\circ),\ (60^\circ,\ 60^\circ)$ no approximately commensurate boundaries with length less than 2000 \AA\ could be found, and thus no boundaries were simulated at these grid points. Fig.~\ref{fig:gbspace}B shows $(\theta_{\rm{M}},\ \theta_{\rm{L}})$ for all GBs configurations that we simulated (4122 total). Each point in that figure represents several simulations due to the sampling of the relative sliding $\Delta$. In all we have simulated over 79,000 GB structures for this study.

\subsection{Numerical GB Structure Generation Algorithm}

Experimental observations of well annealed graphene GBs in the present study, as well as by several previous authors,~\cite{huang2011grains, an2011domain, kim2011grain, kaiser2012, kurasch2012atom, lee2013,rasool2013measurement, rasool2014conserved} suggests that the orientation change between adjacent grains meeting at a GB is mediated largely by pairs of rings of 5 and 7 carbon atoms. These pentagon-heptagon pairs, also called the 5-7 pairs, are the dislocation cores with the shortest Burgers vectors in graphene, and have a low formation energy ~\cite{yazyev2010topological}. Thus, it is reasonable to expect that graphene GBs simulated on a computer have atomic structures where the orientation change between the grains is mediated mostly by the experimentally observed pentagon-heptagon pairs. However, it is difficult to meet this requirement in practice. While a few simple GBs composed solely of pentagon-heptagon pairs have been simulated successfully~\cite{grantab2010, zhang2012, yakobson2013, liu2012, hao2012}, deviations from this motif are evident in the GB and polycrystals used in several recent studies~\cite{shenoy2014, gao2012, yannik2012, jannik2013, cao2013}. The reason for this limitation is that so far no computationally efficient method has been proposed to generate well-annealed graphene GBs on a computer. Methods based on grand canonical Monte Carlo simulations, while theoretically sound and simple to implement, take inordinately large amount of computer time in practice. The common method of simply ``annealing'' a grain boundary by running molecular dynamics at an elevated temperature is also not effective since the typical thermal barriers that need to be overcome for suitable reconstruction are high, and thus the desired annealing does not occur in the limited simulation time. For simple GBs these limitations can be overcome by the inclination-disclination based geometric method~\cite{yazyev2010topological}. This geometrical approach is well suited for the study of simple GBs, but becomes unwieldy for tailoring more complex GBs or polycrystals with several different GBs and junctions.

\begin{figure}[htbp]
\begin{center}
\includegraphics[width=3.4in]{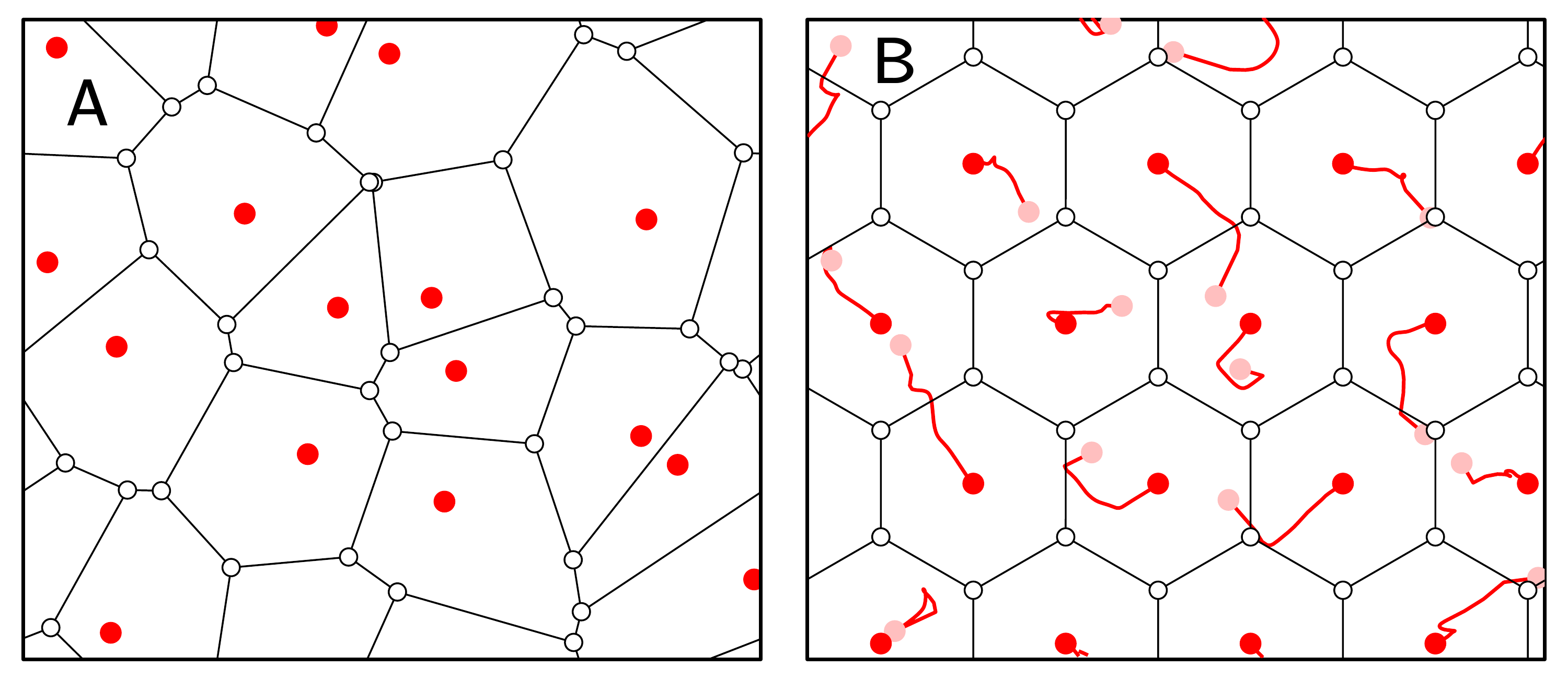}
\caption{(Color online) {\bf (A)} A 2D-periodic Voronoi tessellation with $n=16$ generators. The generators are shown in red circles, the vertices of the Voronoi tessellation obtained from the generators in open black circles, and the edges joining the vertices in solid lines. {\bf (B)} A centroidal Voronoi tessellation (CVT) obtained by applying Lloyd's algorithm to iteratively update the generator positions shown in (A). The dark red circles show the final positions of the generators, while the initial position is shown by the light pink circles. The path traced by the generators during the iterations is shown by the red lines. Note that some paths cross the periodic boundaries. The open black circles show the vertices of the final CVT. }
\label{fig:VoronoiExample}
\end{center}
\end{figure}

To create physically realistic graphene GBs with dislocation density as close as possible to the geometrically required density, we propose an algorithm based on the centroidal Voronoi tessellation (CVT). Before describing the algorithm in detail, we give an intuitive explanation. A triangular lattice can be associated to the graphene lattice via a Voronoi construction (also known as the Dirichlet or Weigner-Seitz construction, or the dual construction), and vice-versa. For example, in Fig.~\ref{fig:VoronoiExample}B the graphene lattice (open black circles) forms the vertices of the Voronoi cells of the triangular lattice (dark red circles), and vice-versa. As discussed earlier, it is difficult to anneal a graphene GB by using classical Monte Carlo methods. If however, one could find a method to anneal the associated triangular lattice, then the graphene GB could be easily recovered from the well annealed triangular lattice by applying the Voronoi construction. Notice that annealing the triangular lattice by using Monte Carlo or MD will be almost  as difficult as annealing the original graphene lattice with similar techniques. The interesting part of our algorithm uses the CVT to efficiently anneal the triangular lattice, and a well annealed graphene GB is recovered from it via a Voronoi construction.

We give a brief introduction to CVTs; details can be found in any number of references including Refs.~\cite{gunzburger1999,liu2009}. Let $\mathbf{X} = (\mathbf{x}_i)_{i=1}^n$, be a set of $n$ points in a compact connected region $\Omega \subset \mathbb{R}^2$. (the generalization to $\mathbb{R}^N$ is analogous).  The points $\mathbf{x}_i$ will be called the generators of the Voronoi tessellation. The Voronoi region $\Omega_i$ corresponding to the generator $\mathbf{x}_i$ is defined as the set of all points that are closer (or equidistant) to it than to any other generator, i.e., $\Omega_i = \{ \mathbf{x} \in \Omega\ \big|\ \| \mathbf{x} - \mathbf{x}_i\| \leq \| \mathbf{x} - \mathbf{x}_j\|,\ \forall j \neq i \},$ where $\|\cdot\|$ is the usual Euclidean norm. We denote the set of the vertices of the Voronoi regions by $\mathbf{v}_i$. Fig.~\ref{fig:VoronoiExample}A shows an example of a 2D periodic Voronoi tessellation with $n=16$ generators. Clearly, the centroid of the region $\Omega_i$ is in general distinct from its generator $\mathbf{x}_i$. If we demand that the generators be arranged so that the centroids of the resulting regions coincide with their generators, then we get a CVT. A CVT can also be described in terms of a variational problem~\cite{liu2009}. It has been noted that the generators of a CVT are local or global minimizers of the following energy function
\begin{equation}
\mathbf{H}_{CVT}(\mathbf{X}) = \sum_{i=1}^{n} \int_{\Omega_i} \| \mathbf{x} - \mathbf{x}_i \| d\mathbf{x}.
\label{eq:cvtHamiltonian}
\end{equation}
Fig.~\ref{fig:VoronoiExample}B shows an example of a 2D-periodic centroidal Voronoi tessellation with $n=16$ generators. Note that the vertices of the Voronoi regions form a graphene-like hexagonal lattice, while the generators form a triangular lattice. In fact, it is a general property of CVTs that they tend to generate a tessellation with regular hexagonal regions of equal size~\cite{gunzburger1999,liu2009}. The tessellation shown in Fig.\ \ref{fig:VoronoiExample}B is obtained by starting from the configuration of generators shown in \ref{fig:VoronoiExample}A and moving them according to Lloyd's algorithm so as to minimize the energy function $\mathbf{H}_{CVT}(\mathbf{X})$~\cite{gunzburger1999,liu2009}. The path traced by each generator under the action of this algorithm is shown by the red lines in Fig.~\ref{fig:VoronoiExample}B. We choose the aspect ratio of the domain such that it is possible to tile it with 16 regular hexagons. The tessellation is 2D-periodic because we implement 2D-periodic boundary conditions in our metric $\| \cdot \|$. A perfect tessellation with equal regular hexagons does not always exist, and neither does Lloyd's algorithm converge to it from every possible initial condition even if it exists. For example, if we take $n = 15$ generators, then a perfect tessellation is impossible. In such cases, CVTs try to minimize the deviation from perfect hexagons, and on most occasions find a tessellation containing suitable pentagons-heptagons defects.

\begin{figure}[htbp]
\begin{center}
\includegraphics[width=2.4in]{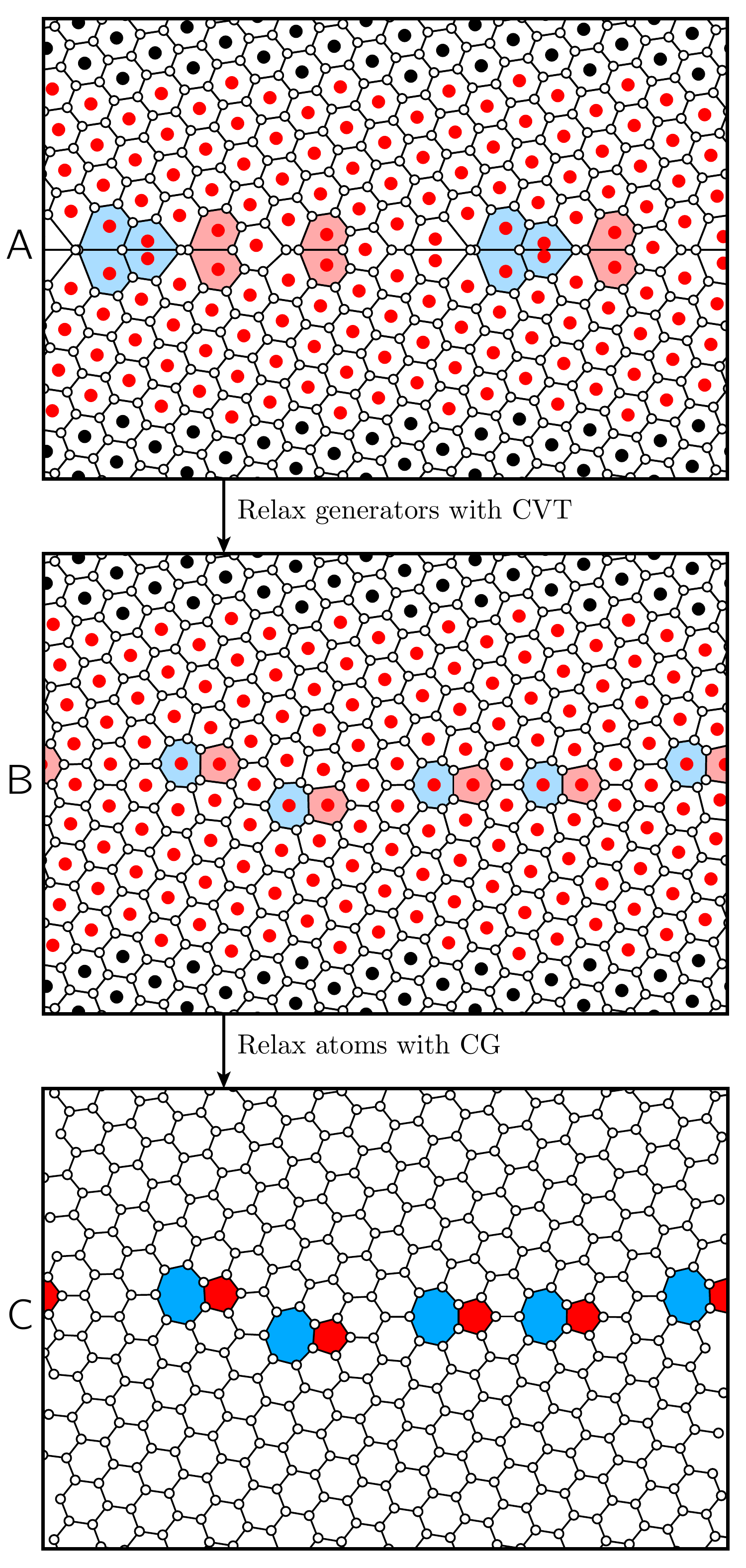}
\end{center}
\caption{
{\bf (A)} A triangular lattice with the appropriate orientation is generated on each side of the GB. The triangular lattice points within 10 \AA\ of the GB shown in red dots, while those further away are shown in black dots. Vertices of the Voronoi tessellation of these points make graphene-like hexagons in the grain interiors, but not near the boundaries, as seen in the figure. 
{\bf (B)} The points within 10 \AA\ of the GB are relaxed using Lloyd's algorithm in order to obtain a CVT. The vertices of the Voronoi tessellation of the relaxed lattice now comprise of hexagons and pentagon-heptagon pairs near the GB. The defects (pentagon-heptagon pairs) are colored for clarity.
{\bf (C)} The GB configuration obtained by placing a carbon atom at each vertex of Voronoi regions is relaxed further by using the AIREBO potential and the conjugate gradient algorithm. The triangular lattice points are no longer shown for clarity.}
\label{fig:VoronoiConstruction}
\end{figure}

The CVT based algorithm for generating graphene GBs is as follows. Given a GB the goal is to decide the positions of carbon atoms so that on each side of the GB the graphene sheet has the desired orientation, while the GB is comprised mostly of pentagon-heptagon dislocations (or undefected hexagons). To achieve this goal, the algorithm first generates a triangular lattice dual to the graphene lattice with suitable orientation on each side of the GB. Fig.~\ref{fig:VoronoiConstruction}A shows an example of this construction. At this point, the Voronoi regions (with the triangular lattice points as generators) contain suitably aligned hexagons away from the GB, but near the boundary the structure is not composed of well-aligned pentagon-heptagon pairs, as shown in the figure. The generators (triangular lattice points) close to the grain boundaries are then relaxed by using Lloyd's algorithm to obtain a CVT while keeping the points that are sufficiently far away from the boundaries fixed. The fixed points are shown by black circles, while the points that are relaxed by Lloyd's algorithm are shown in red circles in Fig.~\ref{fig:VoronoiConstruction}. After the relaxation has converged, we obtain a 2D-periodic Voronoi tessellation for the entire lattice, and obtain a graphene GB by placing a carbon atom at each vertex of the tessellation. Fig.~\ref{fig:VoronoiConstruction}B shows the graphene GB corresponding to the grain structure of Fig.~\ref{fig:VoronoiConstruction}A obtained after this relaxation. We see that the grain interiors are defect free and have the desired orientations, while the GB is mediated by well-aligned pentagon-heptagon pairs. Finally, the atomic positions can be fine tuned by using the congugate gradients method and an atomistic potential; we use the AIREBO potential in this study~\cite{stuart2002}. The graphene GB obtained after this fine tuning is shown in Fig.~\ref{fig:VoronoiConstruction}C. This final step only leads to small changes in the atomic positions, and does not entail any larger topological rearrangements of rings and defects. The numerical implementation is efficient, and we are able to obtain a well annealed GBs that are hundreds of nanometers long in a matter of minutes on a laptop.

\begin{figure}[htbp]
\begin{center}
\includegraphics[width=3.4in]{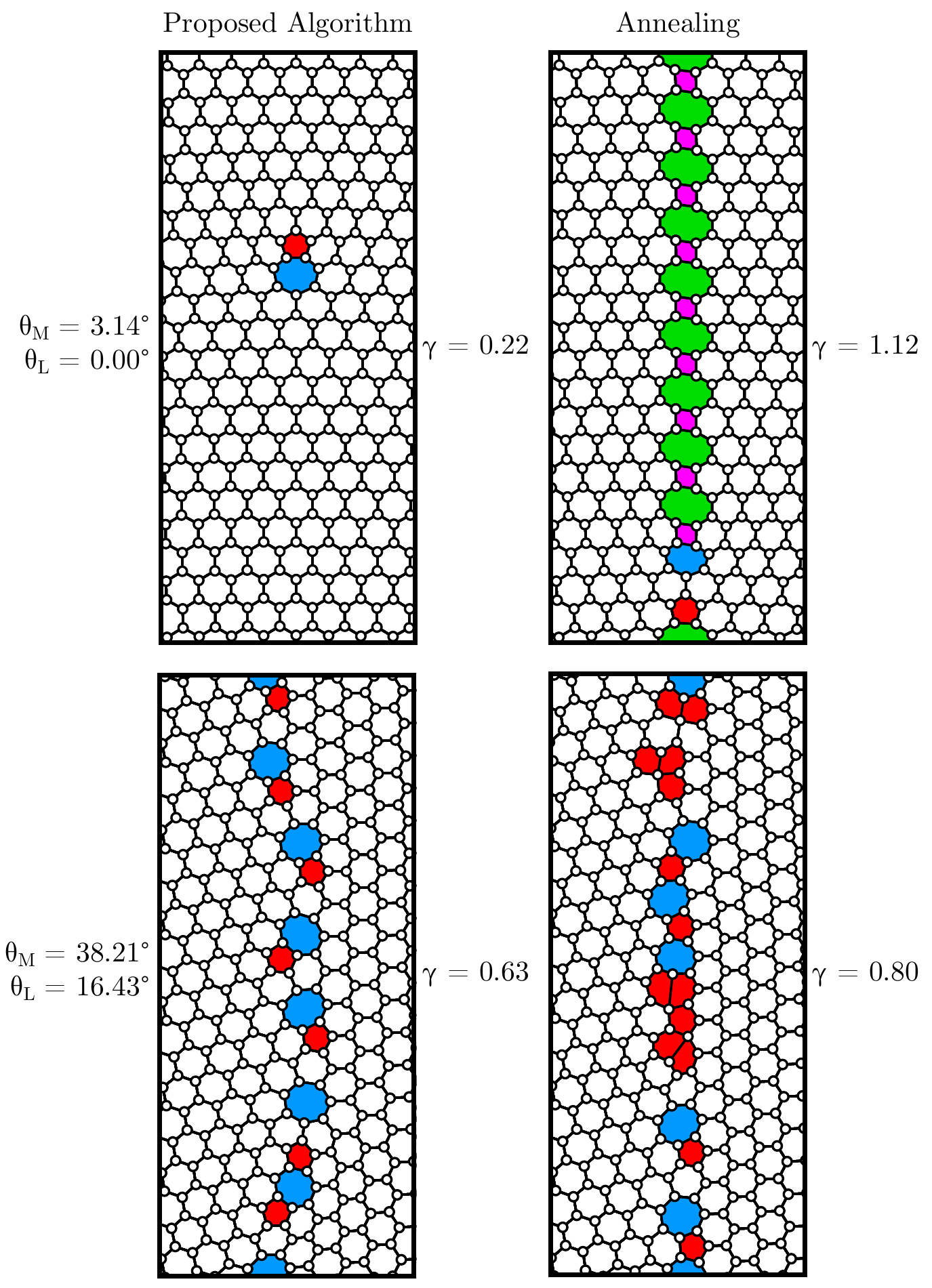}
\end{center}
\caption{A comparison of the GB structures and energies $\gamma$ (in eV/\AA) generated by the proposed algorithm and the widely used technique of annealing. For both the examples shown here the GB generated with the proposed algorithm has significantly lower energy, and is thus closer to the true ground state than those produced with the annealing process. This behavior is generic, and in all the cases that we have evaluated the proposed method almost always results in lower GB energy.}
\label{fig:VoronoiComparison}
\end{figure}

We have compared the GB structures generated with the proposed algorithm with the other widely used methods of generating GBs. One popular technique is to paste together two half crystals of the required orientations and anneal the system by running molecular dynamics at an elevated temperature~\cite{shenoy2014}. We use this technique, where we heat the GB from 10 K to 3000 K, and then cool it back to 10 K in a span on 100 ps. The final configuration is then relaxed by using the conjugate gradient method. During this entire process a 10 \AA\ strip of atoms on the left and right edges of the system are constrained to their ideal crystalline positions. The net width of the system excluding the constrained atoms is 50 \AA. Fig.~\ref{fig:VoronoiComparison} shows the comparison of two GB structures obtained with this method to those obtained by the CVT based method proposed here. The GBs have identical number of atoms, and identical atomic positions away from the boundary. We evaluate the GB energy per-unit length, defined as $\gamma = (E_\mathrm{total} - n_\mathrm{atoms}*E_\mathrm{bulk})/l_{GB}$, where $E_\mathrm{total}$ is the net potential energy of the unconstrained atoms, and $E_\mathrm{bulk} = -7.807$ eV is the ground state energy per atom in graphene according to the AIREBO potential. It is evident that for the examples shown in Fig.~\ref{fig:VoronoiComparison} proposed method outperforms the method of annealing as it generates GBs with lower energies. We have tested several hundred GBs, and the proposed method always performs better than the method of annealing. 


To understand why this boundary generation algorithm outperforms the traditional method of annealing or grand canonical Monte Carlo or simple energy minimization, we note that the energy landscape of the CVT Hamiltonian $\mathbf{H}_{CVT}(\mathbf{X})$ is in some sense more favorable than that of the conventional atomistic potential based Hamiltonian. While just like the atomistic potentials, the CVT Hamiltonian can have several local minima, it seems that unlike the atomistic potentials, all its local minima represent low energy configurations of the polycrystalline graphene sheet. In a perfect tessellation, each generator contributes two vertices, thus removing or adding a generator is analogous to creating a vacancy pair or an adatom pair. This is a very desirable property, since it ensures that isolated vacancies or adatoms never appear, as these are high energy defects~\cite{arkady2011}. All the vacancy pair and adatom pair defects generated by removing and adding generators are shown in the supplemental material and correspond to low energy configurations of vacancy and adatom pairs. Thus, the algorithm is able to produce realistic grain boundaries as well as point defects.

The CVT Hamiltonian is oblivious of all the nuanced and complicated interactions between carbon atoms, as it takes a geometric view of the problem. This is a strength and a weakness of this approach. Its strength is clearly demonstrated in the high quality structures that it can generate at modest numerical cost. Its weakness would be that it is hard, if not impossible, to modify this approach to include, say, the effect of chemical interactions with hydrogen (or another element) on structure of the GB. However, since the structure and properties of pure graphene GBs and similar two-dimensional materials are of such wide interest, we think that the proposed method has broad merit. Finally, it should be noted that the primary role of the CVT algorithm is to relax the triangular lattice. As mentioned before, it is not feasible to simply use a LJ potential (or another pair potential, or hard spheres etc.) to relax this triangular lattice and simplify this algorithm. Thus, the unique properties of the CVT truly offer an advantage over pair potentials and Monte Carlo based methods in this case.




\section{Results and Discussion}

The full library of our measured experimental GB structures is available at the \href{https://sites.google.com/a/lbl.gov/atomic-structure-repository/graphene-grain-boundaries---experiment}{experimental structures archive}.  The full library of our simulated GB structures is available at the \href{https://sites.google.com/a/lbl.gov/atomic-structure-repository/graphene-grain-boundaries---computed}{computed structures archive}.

\begin{figure*}[htbp]
    \centering
        \includegraphics[width=6.4in]{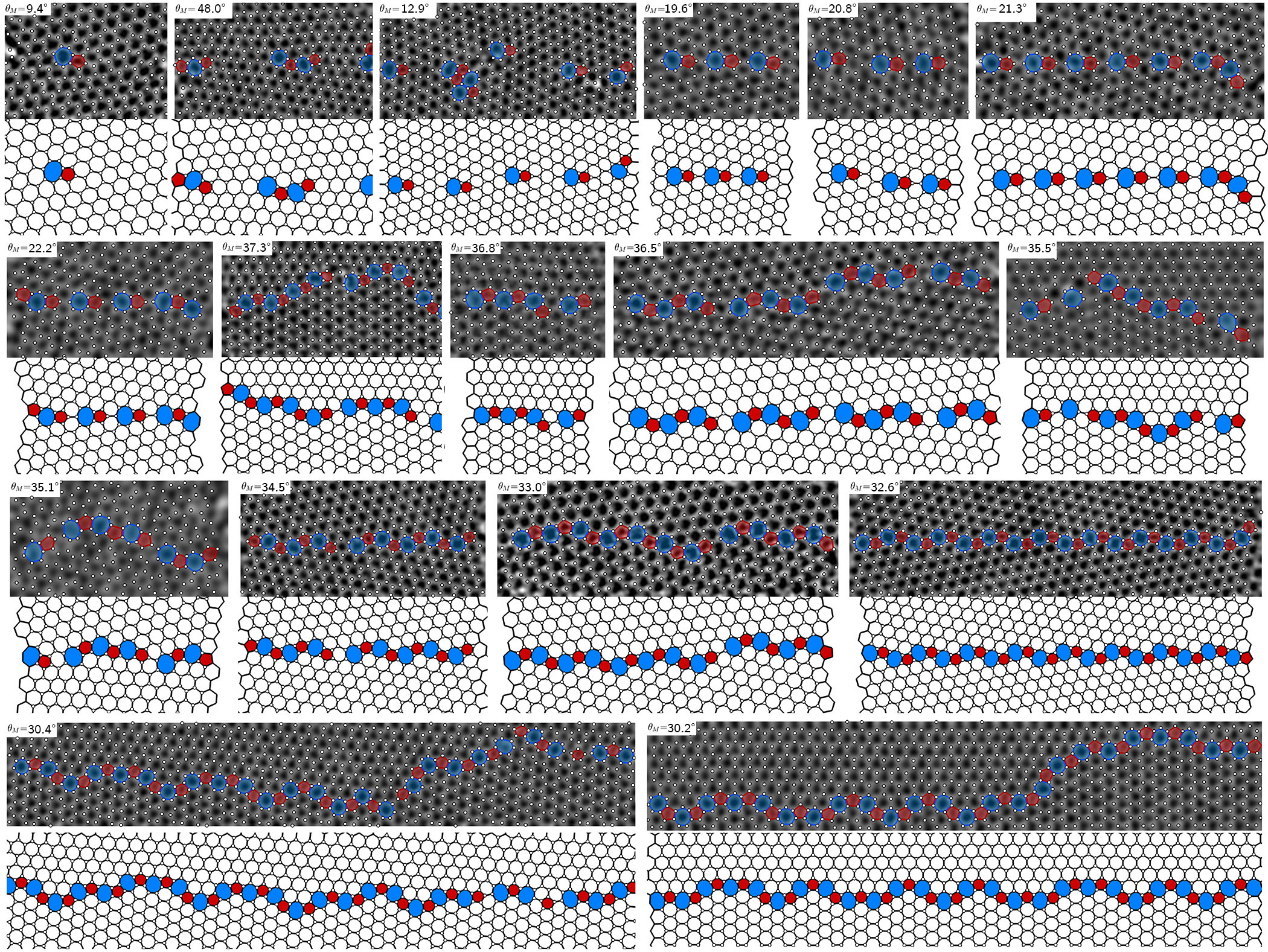}
 	\caption{Examples of experimentally measured graphene GBs (above) compared to similar numerically simulated boundary structures (below), sorted by disorientation angle $\| \left( \theta_{\rm{M}} + {30^{\circ}} \! \mod  60^{\circ} \right) - 30^{\circ} \|$.  Boundary structures range from isolated dislocations, periodic arrays of separated dislocations, continuous high-symmetry boundaries to serpentine boundaries with large amounts of excess dislocation content.}
	\label{FigureExpMDCompare}
\end{figure*}

Fig.~\ref{FigureExpMDCompare} shows 17 examples of experimentally measured GB structures, ranging from low to high boundary disorientations $\| \theta_{\rm{M}} \|$. These misorientations are calculated from the best-fit lattices of the two grains, with an estimated error of approximately 0.5$^{\circ}$. The low angle boundaries are composed of isolated pentagon-heptagon pairs, while the higher angle boundaries are composed of connected pentagon-heptagon pairs. Each experimental boundary is paired with a matching example taken from the generated boundary library, with either an identical or very similar structure. The close agreement demonstrates the efficacy of our boundary generation algorithm.

\subsection{Simulated Grain Boundary Structures}

\begin{figure}[!ht]
    \centering
        \includegraphics[width=2.6in]{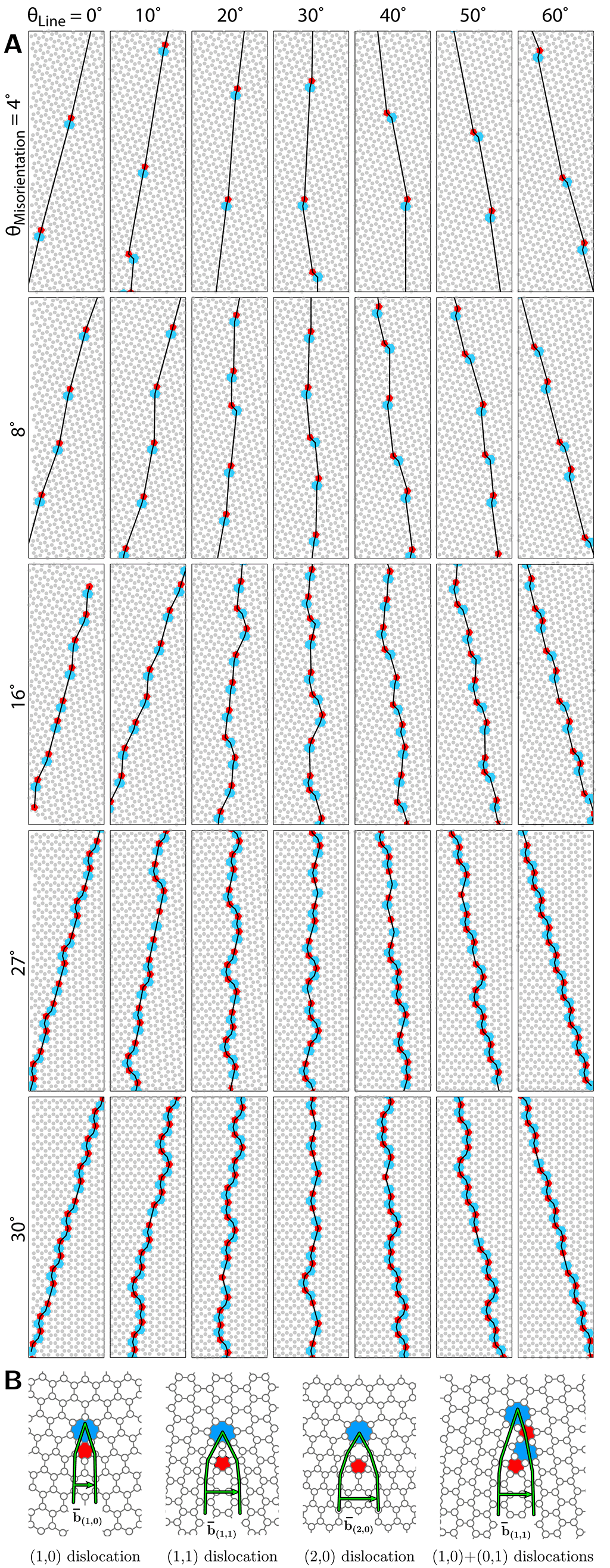}
 	\caption{{\bf(A)} Examples of the GBs calculated with our boundary generation algorithm for various misorientation angles $\theta_{\rm{M}}$ and boundary line angles $\theta_{\rm{L}}$. {\bf(B)} Dislocation structures present in low-energy graphene GBs.}
	\label{FigureMDStructures}
\end{figure}

A small subset of the numerically simulated GB structures are plotted in Fig.\ \ref{FigureMDStructures}A. The 5-member pentagon rings are colored in red, while the 7-member heptagon rings are colored in blue. Each pentagon-heptagon pair sharing a C-C bond represents a (1,0) dislocation core with the smallest possible Burgers vector, while a pentagon-heptagon pair connected by a C-C represents a (1,1) dislocation core with the next-smallest Burgers vector \cite{yazyev2010topological}. Separating the pentagon and hepagon by a single 6-member hexagon ring leads to a (2,0) dislocation with a Burgers vector with twice the magnitude of the (1,0) dislocations. Fig.~\ref{FigureMDStructures}B shows the atomic structure of these dislocations graphically.

Fig.~\ref{FigureMDStructures}B also shows another commonly observed dislocation structure; pairs of (1,0) and (0,1) dislocations.  These dislocation pairs have the same Burgers circuit as the (1,1) dislocation. These dislocation pairs are typically much lower energy than (1,1) dislocations and are commonly observed in the range $ 21.8^{\circ} < \theta_{\rm{M}} < 60^{\circ} $, and are especially prevalent in for misorientation angles $\theta_{\rm{M}} > 32.2^{\circ}$ \cite{yazyev2010topological}.

\subsection{Structural Properties of Simulated Boundaries}

\begin{figure*}[htbp]
    \centering
        \includegraphics[width=6.8in]{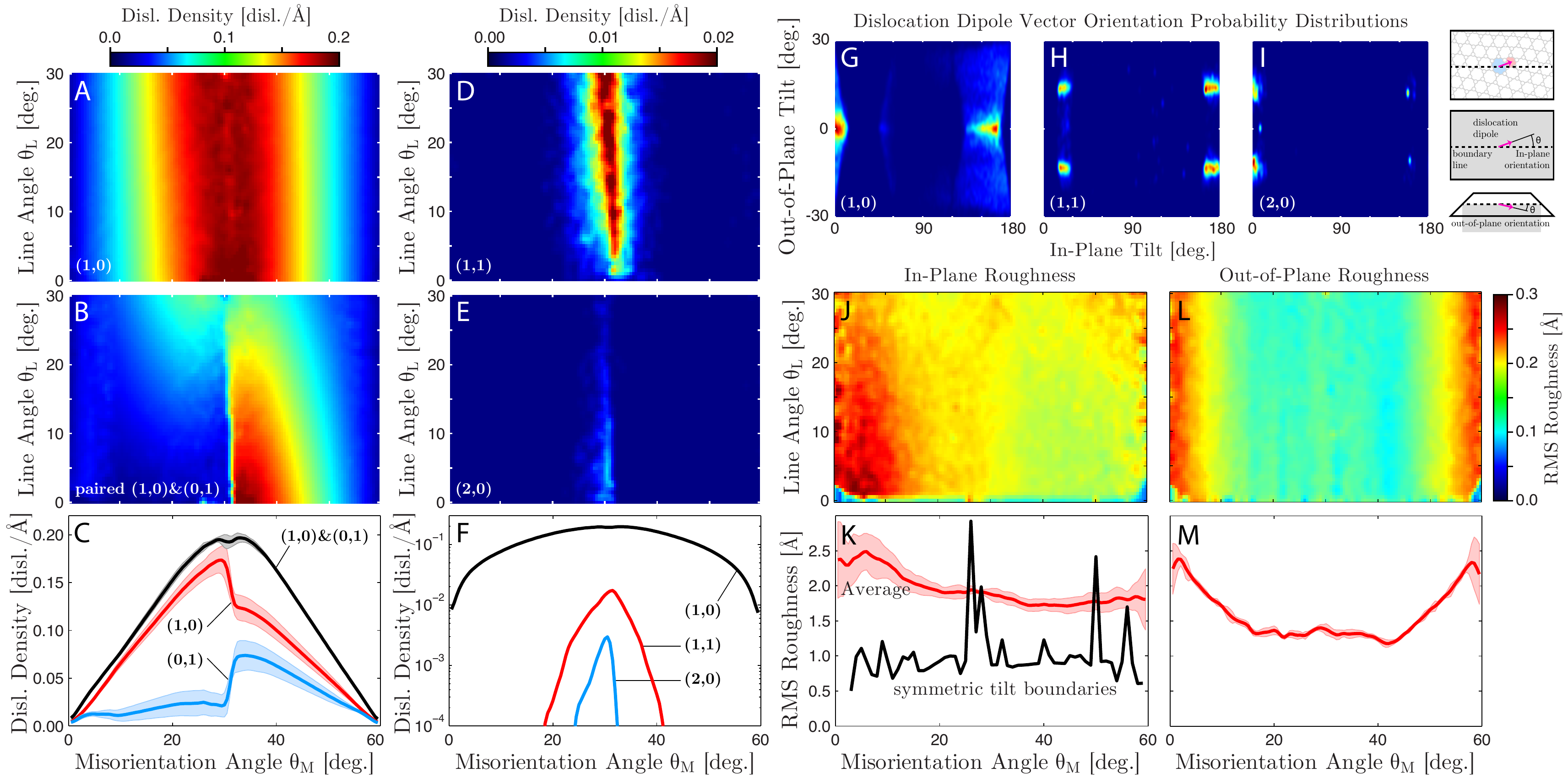}
 	\caption{Dislocation density probability distributions for {\bf(A)} (1,0), {\bf(B)} paired (1,0)-(0,1), {\bf(D)} (1,1) and {\bf(E)} (2,0) dislocations. 
 	{\bf(F)} Average dislocation density as a function of the misorientation angle $\theta_{\rm{M}}$ for (1,0)-type dislocations showing fractions of primary (1,0) and secondary (0,1) orientations. {\bf(F)} Logarithmic probability of average dislocation densities for all types. Dislocation dipole vector orientation probability distributions for {\bf(G)} (1,0), {\bf(H)} (1,1) and {\bf(I)} (2,0) dislocations. Boundary line RMS roughness for {\bf(J)-(K)} in-plane and {\bf(L)-(M)} out-of-plane displacements. Average roughness is plotted as red line, while the standard deviation is plotted as a pink boundary.}
	\label{FigureMDstats}
\end{figure*}

We have used automated analysis routines to extract the dislocation content and structural properties from all of the lowest-energy simulated boundaries for each value of $\theta_{\rm{M}}$ and $\theta_{\rm{L}}$. Figs.~\ref{FigureMDstats}A-F plot the dislocation content of the simulated boundaries. The most common boundary structures by a large margin (over 98\%) are the (1,0) and paired (1,0)+(0,1) dislocations. Figs.~\ref{FigureMDstats}B and C show that the pairing arrangement is much more common for boundaries with $\theta_{\rm{M}} \geq 30^\circ$, although many pairs are also present for the most asymmetric boundaries (high $\theta_{\rm{L}}$).

Figs.~\ref{FigureMDstats}D and E plot the density of (1,1) and (2,0) dislocations, both of which are almost entirely present only in boundaries with high disorientations, $25^\circ < \theta_{\rm{M}} < 35^\circ$. The peak density of (1,1) and (2,0) dislocations is approximately 10 and 60 times lower than the (1,0) dislocation density respectively, shown in Fig.~\ref{FigureMDstats}F. (1,1) dislocations are slightly more prevalent at higher $\theta_{\rm{L}}$ values, while (2,0) dislocations have higher density at lower $\theta_{\rm{L}}$ values.

We have also analyzed the three-dimensional orientation densities of the dislocation dipole vectors, defined as the vector from the center of each heptagon to its associated pentagon. The 2D probability distributions of all dislocations (equally weighted for each calculated boundary) over in-plane and out-of-plane dipole tilt vectors are plotted in Figs.~\ref{FigureMDstats}G, H and I for (1,0), (1,1) and (2,0) dislocations respectively. The (1,0) dislocations tend to align along the boundary line, with two large clusters visible in Fig.~\ref{FigureMDstats}G; the left cluster is formed from the lower $\theta_{\rm{M}}$ boundaries while the cluster to the right contains more high $\theta_{\rm{M}}$ boundaries. These right-side (1,0)-type dislocations tend to be slightly tilted away from the boundary line vector and are often paired with a (0,1)-type dislocation, giving a longer tail towards lower $\theta_{\rm{M}}$ values in this cluster. A third, very dim cluster is visible at approximately $60^\circ$ in-plane tilt values.  All three of these clusters are centered on $0^\circ$ out-of-plane tilt, with the distributions decreasing quickly at higher and lower out-of-plane tilt values.  All three clusters have a range of approximately $\pm 30^\circ$ for the out-of-plane tilt.  By contrast, Fig.~\ref{FigureMDstats}H shows that the (1,1)-type dislocations have a strongly bimodal probability distribution for both in-plane and out-of-plane tilts, occur primarily at in-plane tilts of $\approx$$20^\circ$ and $\approx$$170^\circ$ and out-of-plane tilts of $\pm 13.5^\circ$. The (2,0)-type dislocation orientations are plotted in Fig.~\ref{FigureMDstats}I, showing maxima at an in-plane tilt of $0^\circ$ and out-of-plane tilts of $\pm 12.5^\circ$.

The roughness of all simulated boundaries was estimated by connecting all boundary pentagons and heptagons sequentially, and measuring the root-mean-square (RMS) displacement of this distorted boundary line, both in-plane and out-of-plane of the graphene sheet.  The in-plane RMS roughness is plotted in Fig.~\ref{FigureMDstats}J as a function of the boundary angles.  The mean and standard deviation of the roughness
averaged over all $\theta_{\rm{L}}$ values is plotted in Fig.~\ref{FigureMDstats}K.  The in-plane boundary roughness is largest at low $\theta_{\rm{M}}$ values, decreasing from approximately $2.5\rm{\AA}$ to $2\rm{\AA}$ with increasing $\theta_{\rm{M}}$. The in-plane
roughness of the symmetric tilt boundaries are also plotted in Fig.~\ref{FigureMDstats}. Most symmetric tilt boundaries have lower roughness than the average of all boundaries,  approximately 1\AA, except for a small number of boundaries spiking at RMS roughness values of 3\AA.  The out-of-plane RMS roughness of all boundaries is plotted in
Figs.~\ref{FigureMDstats}L and M.  These roughness values are much more uniform that the in-plane roughness; between $\theta_{\rm{M}} = 0^\circ$ and $15^\circ$ the roughness decreases from $2.4\rm{\AA}$ to 1.3\AA. From $\theta_{\rm{M}} = 15^\circ$ and $45^\circ$ the roughness is almost constant at 1.3\AA.  Finally, between $\theta_{\rm{M}} = 45^\circ$
and $60^\circ$ the roughness increases from $1.3\rm{\AA}$ to 2.4\AA. The symmetric boundaries do not show any deviation in out-of-plane roughness compared to the average values.  The boundaries with larger disorientations have higher dislocation densities;
this allows their overlapping strain fields to more easily cancel out and therefore lead to lower out-of-plane roughness.

\subsection{Structural Properties of Experimental Boundaries}

\begin{figure}[htbp]
    \centering
        \includegraphics[width=3.3in]{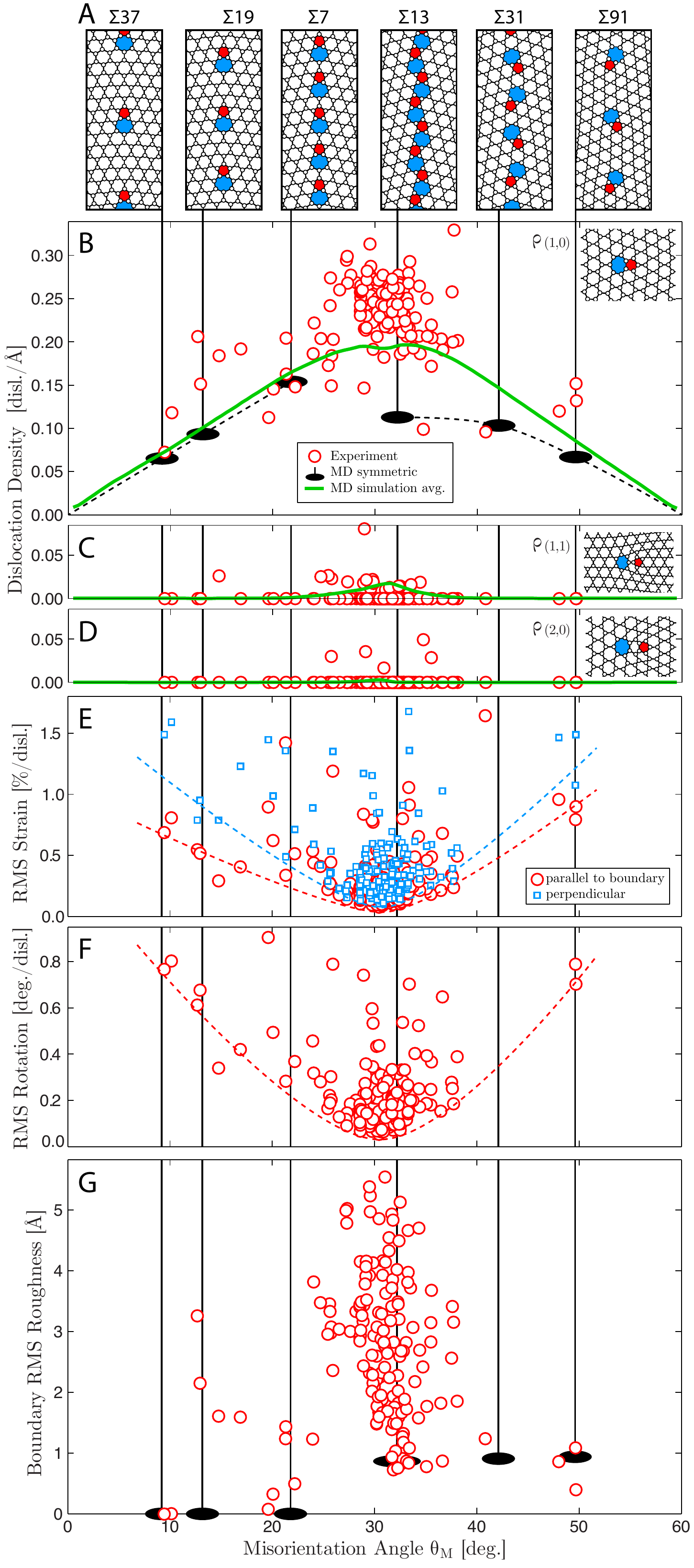}
 	\caption{  {\bf(A)} Examples of special symmetric boundaries with the smallest repeat lengths. Experimental dislocation density for {\bf(B)} (1,0), {\bf(C)} (1,1) and {\bf(D)} (2,0) dislocations respectively.
 	{\bf(E)} RMS atomic strain of experimental boundaries, parallel and perpendicular to boundary line. {\bf(F)} RMS atomic rotations of experimental boundaries. Trendlines shown for experimental strain and rotation.
 	{\bf(G)} In-plane RMS roughness of experimental boundaries and the  boundaries in {\bf(A)}.}
	\label{FigureExpstats}
\end{figure}

The physical properties and structure of the experimental boundaries were also characterized with automated routines.  These results are shown in Fig.~\ref{FigureExpstats}. Some examples of symmetric tilt boundaries with structures similar to those plotted in Fig.~\ref{FigureExpMDCompare} are depicted in Fig.~\ref{FigureExpstats}A. The experimental results plotted in the rest of Fig.~\ref{FigureExpstats} are 160 boundaries estimated to be unique structures. We took this step to try to minimize double counting of boundary structure datapoints.  We observe that most of the boundary structures we measured fall at high disorientation angles, i.e. boundaries close to $\theta_{\rm{M}}=30^{\circ}$.  This phenomenon is due to topological effects; low angle boundaries have rougher surfaces and long range out-of-plane distortions~\cite{liu2010cones}.  This topology in turn attracts carbon contamination due to surface charging, which obscures the boundary.  The imaging process is therefore biased towards flatter boundaries, i.e. those closer to $\theta_{\rm{M}}=30^{\circ}$.

Figs.~\ref{FigureExpstats}B, C and D shows the measured densities as a function of misorientation angle $\theta{\rm{M}}$ for (1,0), (1,1), and (2,0) dislocations respectively.  These figures also show the (1,0) dislocation densities of the 6 symmetric boundaries plotted in Fig.~\ref{FigureExpstats}, and the dislocation densities of all three types predicted from the simulated boundary relaxations in Fig.~\ref{FigureMDstats}F. The experiments are in good agreement with both of these sets of predictions. All of the 6 symmetric boundaries shown in Fig.~\ref{FigureExpstats}A have a nearby experimental example.  However at misorientation angles in the range $25^\circ < \theta_{\rm{M}} < 35^\circ$, in the highest density region of the experimental boundaries, the average predictions of the relaxed and constructed boundaries are much closer to the majority of experimental dislocation densities. The simulated boundaries also predict a small concentration of (1,1)- and (2,0)-type dislocations in the range $25^\circ < \theta_{\rm{M}} < 35^\circ$, both of which are observed in the experimental measurements shown in Figs.~\ref{FigureExpstats}C and D.  The average dislocation concentrations of the experiments are very close to the simulations, with the exception of a single (1,1) dislocation observed in an experiment at $\theta_{\rm{M}}=14^\circ$.

Because the experimental boundaries are measured as a 2D projection, the out-of-plane distortions cannot be directly measured.  However, these distortions are typically accompanied by large deviations in the local projected atomic positions.  We have therefore measured the average strains and local rotations of all boundaries, plotting in Figs.~\ref{FigureExpstats}E and F, normalized by the dislocation density. All of the average strain metrics have approximately the same trend; they decrease as the misorientation increases, and then the strain increases past $\theta_{\rm{M}} = 30^\circ$.  This is qualitatively in agreement with the predictions of out-of-plane roughness trends shown in Fig.~\ref{FigureMDstats}M. The RMS strain perpendicular to the grain boundaries is large than the parallel strain for virtually all boundaries, shown in Fig.~\ref{FigureExpstats}E. This is because dislocation dipoles are typically aligned along the grain boundaries, which allows the adjacent dislocation strain fields to partially cancel out.

The boundary RMS roughness for all experimental boundaries is plotted in Fig.~\ref{FigureExpstats}F. Five of the six symmetric boundaries plotted in Fig. \ref{FigureExpstats}A predict boundary roughness values that are very close matches to the experiments. The largest concentration of boundaries near the center of the plot reach a minimum roughness at a misorientation value closer to the $\Sigma13$ value of $\theta_{\rm{M}} = 32.2^\circ$. The entire cluster of values has an average in-plane RMS roughness of approximately 2.5\AA, in good agreement with the synthetic boundary prediction of 2\AA given in Fig.~\ref{FigureMDstats}K.

\subsection{Matching Experimental and Simulated Structures}

The vast majority of experimentally measured GBs and all of the numerically simulated GBs in this study can be constructed by mixing the dislocations structures shown in Fig.~\ref{FigureMDStructures}B. Symmetric boundaries with $\theta_{\rm{M}} \leq 21.8^\circ$ contain aligned (1,0) dislocations, while non-symmetric boundaries and boundaries with $\theta_{\rm{M}} > 21.8^\circ$ are typically composed of a mixture of (1,0) and (0,1) dislocations~\cite{yazyev2010topological}. The ratio of the number of the two most common orientations for (1,0) dislocations could be used to estimate the boundary line angle $\theta_{\rm{L}}$, but for many of the experimental images the boundary length is too short (not enough observed dislocations) for an accurate measurement of the line angle.

As predicted, the low angle boundaries consist of isolated (1,0) dislocations. At low misorientation angles, all of the (1,0) dislocations have the same orientation, while at high misorientation angles (e.g.\ $\theta_{\rm{M}}=48^{\circ}$) the structure consists of (1,0) and (0,1) dislocation pairs. Three of the plotted examples in Fig.\ \ref{FigureExpMDCompare}, $\theta_{\rm{M}} = 20.8^{\circ}$, $21.3$, and $22.2$, have structures very similar to the $\Sigma 7$ special boundary~\cite{heckl1992domain, simonis2002stm, yazyev2010topological}. The $\theta_{\rm{M}} = 32.6^{\circ}$ boundary is a nearly perfect example of the $\Sigma 13$ special boundary~\cite{yazyev2010topological}.

The high angle boundaries in Fig.~\ref{FigureExpMDCompare} are formed from continuous or near-continuous dislocation groups with alternating 5- and 7-member rings. This structure is expected, since a more equal local density of pentagons and heptagons leads to a lower net disclination content and thus less 3D topological variation, and more stable structures~\cite{liu2010cones}. The longest high angle boundaries in Fig.~\ref{FigureExpMDCompare} ($\theta_{\rm{M}}=30.4^{\circ}, 30.2^{\circ}$) show an interesting deviation from the flat boundary structures; they form serpentine structures similar to some literature predictions and observations~\cite{huang2011grains, tison2014grain, vancso2014effect} as well as our previous experiments~\cite{rasool2013measurement, rasool2014conserved}.  Both of these boundaries exhibit two relatively sharp $30^{\circ}$ deviations from a flat boundary line, where the arm-chair and zig-zag graphene edges of the two sides exchange identities. These structures likely originate from capillary fluctuations during the initial growth, but form (relatively) well-defined faceted edges rather than a rougher boundary.

Overall, the boundary generation algorithm therefore accurately predicts structural properties of the experimental boundaries.  The primary disagreement is the slightly higher boundary roughness and dislocation density of the high angle boundaries near $\theta_{\rm{M}}$.  The source of this minor disagreement lies in the faceting exhibited by the experimental boundaries, such the bottom two experimental structure plots in Fig.\ref{FigureExpMDCompare}. Since the simulated boundaries were constrained to follow a single line angle (flat boundaries), they could not fully capture this effect.  However, our CVT algorithm could easily be used to simulate such boundaries.  This example shows how large-scale experiment and simulations couple together, where the different areas of agreement or disagreement can be used to improve structure models.


\section{Conclusion}

In summary, we have used semi-automated processing routines to characterize the structure of a very large number of experimentally measured single-layer graphene grain boundaries and described their local atomic structure as a function of misorientation angle.  We have also introduced a new algorithm for generating realistic graphene grain boundaries that produces structures in excellent agreement with the experimental boundary structures.  We have used a combination of our algorithm and molecular dynamics to generate and relax graphene grain boundary structures covering the entire orientation parameter space for single-later graphene boundaries. The structure and physical properties of these simulated boundaries were analyzed as a function of misorientation and boundary line angles. A detailed comparison of the experimental and simulated boundaries demonstrates that our structural models have high predictive power for the experimental structures.  In a forthcoming study, we will analyze the energetics of both experimental and simulated  grain boundaries.  Finally. all experimental and simulated structures are made available on the internet.  We hope that this paradigm of computer-assisted analysis of a statistically relevant number of structures and availability of all measured data becomes standard for the study of atomic-resolution structures.

\section{Acknowledgements}

Work at the Molecular Foundry was supported by the Office of Science, Office of Basic Energy Sciences, of the U.S. Department of Energy under Contract No. DE-AC02-05CH11231. CO thanks Matt Bowers, Ulrich Dahmen and Josh Kacher for useful discussions. AS acknowledges financial support from the Miller Institute for Basic Research in Science, at University of California, Berkeley, in the form of a Miller Research Fellowship, and thanks Robert O. Ritchie for hosting him at the Lawrence Berkeley National Laboratory.

\bibliography{referencesGrapheneGB.bib,references_ashivni.bib}

\end{document}